\documentstyle[12pt]{article}
\catcode`\@=11

\def\@fnsymbol#1{\ifcase#1\or * \or \dagger\or \ddagger\or
\mathchar "278\or \mathchar "27B\or \|\or **\or \dagger\dagger
\or \ddagger\ddagger \else\@ctrerr\fi\relax}%

\def\title{%
\vspace{0.5cm}\vspace{4ex}
\bgroup
\obeylines
\large\boldmath \bf\begin{center}
}
\def\endtitle{\end{center}\vskip1sp\egroup}
\def\author#1{\begingroup\center #1 \endcenter\endgroup}

\def\keywords#1{\par\noindent{\bf KEYWORDS}: #1}
\def\addcontentsline#1#2#3{\relax}
\def\fnum@figure{Figure \thefigure}
\def\fnum@table{Table \thetable}
\newcounter{figcaption}
\def\thefigcaption{\arabic{figcaption}}
\def\fnum@figcaption{{\bf Fig. \thefigcaption :}}
\def\figcaption{%
{\parindent 0pt \bf Figure Captions} \par \vskip 10pt
\list{\fnum@figcaption}
{\leftmargin 5em \labelwidth\leftmargin\advance\labelwidth-\labelsep
\def\makelabel##1{##1\hfil} \usecounter{figcaption}}%
}

\def\cite{\@ifnextchar [{\@tempswatrue\@citex}{\@tempswafalse\@citex[]}}
\def\@citex[#1]#2{\if@filesw\immediate\write\@auxout{\string\citation{#2}}\fi
\def\@citea{}\@cite{\@for\@citeb:=#2\do
{\if-\@citeb \mbox{-}\def\@citea{}
\else
\@citea\def\@citea{,\penalty\@m}\@ifundefined
{b@\@citeb}{{\bf ?}\@warning
{Citation `\@citeb' on page \thepage \space undefined}}%
\hbox{\csname b@\@citeb\endcsname}
\fi}}{#1}}
\newfont{\scrptrm}{cmr8}
\def\@cite#1#2{${}^{\scrptrm {#1\if@tempswa , #2\fi})}$}

\def\rcite{\@ifnextchar [{\@tempswatrue\@rcitex}{\@tempswafalse\@rcitex[]}}
\def\@rcitex[#1]#2{\if@filesw\immediate\write\@auxout{\string\citation{#2}}\fi
\def\@rcitea{}\@rcite{\@for\@rciteb:=#2\do
{\if-\@rciteb \mbox{-}\def\@rcitea{}%
\else
\@rcitea\def\@rcitea{,\penalty\@m}\@ifundefined
{b@\@rciteb}{{\bf ?}\@warning
{Citation `\@rciteb' on page \thepage \space undefined}}%
\hbox{\csname b@\@rciteb\endcsname}%
\fi}}{#1}}
\def\@rcite#1#2{{#1\if@tempswa , #2\fi}}
\def\refcite{\@ifnextchar [{\@tempswatrue\@refcitex}
{\@tempswafalse\@refcitex[]}}
\def\@refcitex[#1]#2{
\if@filesw\immediate\write\@auxout{\string\citation{#2}}\fi
\def\@citea{}\@refcite{\@for\@citeb:=#2\do
{\if-\@citeb -\def\@citea{}
\else
\@citea\def\@citea{,\penalty\@m}\@ifundefined
{b@\@citeb}{{\bf ?}\@warning
{Citation `\@citeb' on page \thepage \space undefined}}%
\hbox{\csname b@\@citeb\endcsname}
\fi}}{#1}}
\def\@refcite#1#2{[{#1\if@tempswa , #2\fi}]}

\newcommand{\rmc}{{\rm c}}
\newcommand{\rmd}{{\rm d}}
\newcommand{\rme}{{\rm e}}
\newcommand{\rmi}{{\rm i}}
\renewcommand{\Re}{{\cal R}\!{\sl e}\,}
\renewcommand{\Im}{{\cal I}\!{\sl m}\,}

\newcommand{\Ham}{{\cal H}}

\newcommand{\dags}{^\dagger}

\def\brakets#1{\langle#1\rangle}

\def\simle{\mathrel{\mathpalette\@versim<}}   
\def\simge{\mathrel{\mathpalette\@versim>}}   
\def\@versim#1#2{\lower2.5pt\vbox{\baselineskip0pt \lineskip-.5pt
   \ialign{$\m@th#1\hfil##\hfil$\crcr#2\crcr\sim\crcr}}}

\newcommand{\bequ}{ \begin{equation} }
\newcommand{\eequ}{ \end{equation} }
\newcommand{\barr}{ \begin{array} }
\newcommand{\earr}{ \end{array} }
\newcommand{\beqarr}{ \begin{eqnarray} }
\newcommand{\eeqarr}{ \end{eqnarray} }

\newcommand{\baralpha}{ \begin{eqnal} \beqarr}
\newcommand{\earalpha}{ \eeqarr \end{eqnal}}

\catcode`\@=12

\def\MFreq{\rmi \omega_n}

\def\vecn{\vec m}

\def\LSMO{La$_{1-x}$Sr$_x$MnO$_3$}

\def\eV{\mbox{eV}}

\begin{document}

\begin{title}
Temperature Dependence of the  Conductivity
in (La,Sr)MnO$_3$
\end{title}

\author{Nobuo {\sc Furukawa}}
\begin{instit}
  Institute for Solid State Physics,\\
  University of Tokyo, Roppongi 7-22-1,\\
  Minato-ku, Tokyo 106
\end{instit}

\begin{abstract}
Using the Kondo lattice model with classical spins
in infinite dimension,
conductivity in
the perovskite-type $3d$ transition-metal oxide (La,Sr)MnO$_3$
is theoretically studied.
Green's functions as well as spontaneous magnetization are
obtained exactly  on the Bethe lattice as a function of temperature.
Conductivity is calculated from the Kubo formula.
Below the Curie temperature,
 resistivity as a function of magnetization is in a good agreement
with the experimental data.
Anomalous behaviour in the temperature dependence of the
 optical conductivity
observed in (La,Sr)MnO$_3$ is also explained.

\end{abstract}

\vfil


\pagebreak

The discovery of the high-$T_\rmc$ superconducting cuprates has
driven a renaissance in the study of $3d$ transition metal oxides.
One of such materials is the
perovskite-type manganese oxides ($R$,$A$)MnO$_3$ with $R$ and $A$
being rare-earth and alkaline-earth ions, respectively.
Under appropriate hole doping,
the system becomes a
ferromagnet which is explained by a
double-exchange mechanism.\cite{Zener51,Anderson55,deGennes60}\ \ {}

One of the most prominent features in this family of materials
is the giant magnetoresistance (MR)
with negative sign.
Resistivity decreases as the magnetic field is applied.
Sharp drop in resistivity is also observed when
temperature is lowered below the Curie temperature $T_{\rm c}$.
In  {\LSMO},\cite{Tokura94,Urushibara9x}\
resistivity at $T \simge T_\rmc$  is scaled as a function of
field-induced magnetization in the form
$\rho/\rho_0 = 1- C (M_{\rm tot}/M_{\rm s})^2$,
where $M_{\rm tot}$ and $M_{\rm s} = 4 \mu_{\rm B}$ are
the total magnetization and the nominal saturation magnetization,
respectively. Here, resistivity  $\rho$ is normalized by its
zero-field value $\rho_0$.
With an appropriate normalization, resistivity below $T_{\rmc}$
 is also scaled in the same form as a function of spontaneous magnetization.

In {\LSMO}, anomalous behavior
 in the optical conductivity $\sigma(\omega)$ is also
observed.\cite{Okimoto95x}\ \
At the paramagnetic phase around room temperature, a gap-like
structure is observed.
Below $T_\rmc$, unconventional temperature
dependence  in $\sigma(\omega)$ which
extends beyond $\omega \sim 2\eV$ is found.
Transfer of the spectral weight from the high-energy part at
$\omega \simge 1\eV$ to the low-energy part $\omega \simle 1\eV$
is observed as temperature is lowered.
This implies that the electronic structure in {\LSMO} changes
in the energy scale  $ \sim 1\eV$ by changing temperature
in the scale $\sim 10^{-2} \eV$.
The result may not be explained from the rigid-band
picture. Effects of strong electron correlation in the energy scale
$\sim 1\eV$ must be taken into account to understand the properties.

{}From the theoretical point of view,
the Kondo lattice model with ferromagnetic coupling
is studied as a canonical model of {\LSMO}.
The Hamiltonian in the classical spin limit is described as
\bequ
  \Ham =
  - t \sum_{<ij>,\sigma}
        \left(  c_{i\sigma}\dags c_{j\sigma} + h.c. \right)
    -J \sum_i \vec {\hat\sigma}_i \cdot \vec m_i,
    \label{HamSinfty}
\eequ
where $ \vec m_i = (m_i{}^x, m_i{}^y, m_i{}^z)$ denote
localized spins with $|\vec m|^2 = 1$, while $\vec{\hat\sigma}_i$
 represent spins of itinerant electrons.
Electrons and localized spins correspond to
Mn $3d$ electrons in $e_{\rm g}$ and $t_{2\rm g}$ orbitals, respectively.
Hund's coupling $J$
is considered to be  larger
than the electron bandwidth in {\LSMO}.
The author has examined
the model in infinite dimension ($D=\infty$)
with a  Lorentzian
density of states (DOS),\cite{Tokura94,Furukawa94,Furukawa95cx}\ {}
where the MR data in {\LSMO} has been well reproduced.

In this paper, we examine the model on a $D=\infty$
 Bethe lattice, where
we see a semi-circular DOS (S-DOS) in the noninteracting system
$N_0(\varepsilon) = (2/\pi W) \sqrt{ 1 - (\varepsilon/W)^2}$.
The bandwidth $W \equiv 1$ is hereafter
 taken as a unit of energy.
This system has  more realistic points
than those  with  Lorentzian DOS (L-DOS):
It is composed from short-range hopping terms, and the
 DOS has band edges.
Ferromagnetic phase exists under doping, and
$T_\rmc$ in {\LSMO} at $0.1 \simle x \simle 0.25$
is reproduced quantitatively
 at $J=4$.\cite{Furukawa95bx}\ \ {}
Here, we study temperature dependence of the conductivity
at $T \simle T_{\rm c}$ and its relation with spontaneous magnetization.

In an infinite-dimensional lattice system,
the model is mapped to a single-cite
model interacting with a dynamical field
$\tilde G_0$.  
Trace over fermion degrees of freedom is evaluated as
\bequ
 \tilde Z_{\rm f}(\vecn) =
 4 \exp \left( \sum_n \log \det \left[
       \frac{(\tilde G_{0}^{-1} + J\vecn\vec{\hat\sigma})}
	    {\MFreq}
    \right] \rme^{\MFreq 0_{+}} \right),
  \label{defZf}
\eequ
so that the  partition function is given by
$ \tilde Z = \int \rmd\Omega \ \tilde Z_{\rm f}(\vecn)$.
Then, Green's function for the single-cite system is obtained
 in the form
\bequ
   \tilde G(\MFreq) =
   \frac{1}{\tilde Z}  \int \rmd\Omega \ \tilde Z_{\rm f}(\vecn)
     \left(  \tilde G_{0}^{-1}(\MFreq) +
                    J \vecn \vec {\hat\sigma} \right)^{-1}.
\eequ
Self-energy and Green's function for the lattice system are given by
$\tilde \Sigma = \tilde G_0^{-1} - \tilde G^{-1}$ and
$  G(\epsilon,\MFreq)  =
      [\MFreq - (\epsilon-\mu) -\tilde  \Sigma ]^{-1}$,
respectively.
The Weiss field $ \tilde G_0$ is
determined self-consistently as
\beqarr
 \tilde G_0{}^{-1} &=&
	\left(
		\int \! \rmd \epsilon \, N_0(\epsilon) G(\epsilon,\MFreq)
	\right)^{-1}
 	 + \tilde \Sigma.
	\label{defSCE}
\eeqarr

The nominal carrier electron number in
{\LSMO} is $n=1-x$. Hereafter we
use the hole notation so that the hole concentration is
expressed by $x$.
Magnetization is calculated as
$   \brakets{\vec m} =
     \int \rmd\Omega \ \vecn \ {\tilde Z_{\rm f}(\vecn)} / {\tilde Z}$.
Conductivity in $D=\infty$ is
calculated from the formula\cite{Moller92,Pruschke93a}
\beqarr
  \sigma(\omega) &=& \sigma_0
  \sum_\sigma \int \rmd \omega' \ I_\sigma(\omega',\omega'+\omega)
   \frac{ f(\omega') - f(\omega'+\omega)}{\omega},
   \label{Optcond}
 \\
   I_\sigma(\omega_1,\omega_2) &=& \int N_0(\epsilon) \rmd \epsilon \
     W^2 A_\sigma(\epsilon,\omega_1) A_\sigma(\epsilon,\omega_2),
\eeqarr
where
$A_\sigma(\epsilon,\omega) = -\Im G_\sigma(\epsilon,\omega+\rmi\eta) / \pi$
and $f$ is the Fermi distribution function.
The constant $\sigma_0$ gives the unit of conductivity.
Integration over $\epsilon$ gives
\beqarr
  I_\sigma(\omega_1,\omega_2) =
   \frac{W^2}{2\pi^2}
  \Re \left(
	\frac{ \tilde G_{\rm R \sigma}(\omega_1)
			- \tilde G_{\rm R \sigma}(\omega_2)}
 { G_{\rm R\sigma}^{-1}(\mu,\omega_1) - G_{\rm R\sigma}^{-1}(\mu,\omega_2)}
	-
	\frac{ \tilde G_{\rm R \sigma}(\omega_1)
			- \tilde G_{\rm A \sigma}(\omega_2)}
 { G_{\rm R\sigma}^{-1}(\mu,\omega_1) - G_{\rm A \sigma}^{-1}(\mu,\omega_2)}
		\right), \qquad
\eeqarr
where $\tilde G_{\rm R}$ and $\tilde G_{\rm A}$ ($G_{\rm R}$ and $G_{\rm A}$)
 are the  retarded and the advanced
Green's function of the single-cite (lattice) system, respectively.
We calculate the resistivity as
$\rho = 1/\sigma(\omega\to0)$.

Let us now examine temperature dependence of the resistivity.
We choose   $J=4$ and $x=0.175$,
where
 $T_\rmc = 0.022$. We calculate the resistivity $\rho(T)$
normalized by the value at $T_\rmc$, $\rho_0 = \rho(T_\rmc)$,
and the total spontaneous magnetization $M_{\rm tot}(T)$
normalized by the nominal saturation magnetization $M_{\rm s}$.
Since we consider $t_{2\rm g}$ electrons with $S=3/2$ and itinerant
$e_{\rm g}$ electrons, we have
$ M_{\rm tot} = \frac32 \brakets{m_z} + \frac12\brakets{\hat\sigma_z}$
 and  $M_{\rm s} = 2$.
Here, $z$-axis is taken to be
parallel to the magnetization.
We also
calculate the resistivity and the total induced magnetization
by applying magnetic field
at a fixed temperature in the paramagnetic region
 $T=1.01T_\rmc$ for comparison.
In this case, we normalize the resistivity $\rho(H)$
by its zero-field value $\rho_0$.

In Fig.~\ref{FigMR},  we show $\rho/\rho_0$
as a function of $M_{\rm tot} / M_{\rm s}$
for the both cases
of calculation by  changing temperature $T$ and
the external magnetic field $H$ at the system with the S-DOS.
We also show the result for
L-DOS in the equal condition, where
the resistivity is changed by the magnetic field.\cite{Tokura94}\ \ {}
The difference between any of the three
 cases are small in the wide region of
$M_{\rm tot} / M_{\rm s}$.
Thus, magnetization is the essential
thermodynamical variable that determines the nature of
the resistivity, since   quasi-particle properties such as
life-time and DOS are  strongly
magnetization dependent.\cite{Furukawa94,Furukawa95cx}\ \ {}
Effects of temperature is subsidiary here, because we are in the
limit $T \ll W$.
It is analogous to the nature in {\LSMO}
that the resistivity is scaled by the magnetization
in a universal way.

We then  make a quantitative comparison
 with the experimental data of {\LSMO} at $x=0.175$
where $T_{\rm c} = 283{\rm K}$.\cite{Urushibara9x}\ \ {}
In Fig.~\ref{FigMR}, we plot the resistivity at $T=294{\rm K}$
as a function of induced magnetization normalized by its zero-field value.
Temperature dependence of the resistivity below $T_\rmc$ with
an appropriate normalization is also
shown as a function of spontaneous magnetization.
The experimental result in {\LSMO}
is reproduced quantitatively  at $M_{\rm tot}/M_{\rm s} \simle 0.2$.

Now, optical conductivity is calculated at $J=2$ and $x =0.2$.
In Fig.~\ref{FigOC}, we show  $\sigma(\omega)$
at  $T=1.05T_\rmc$,  $0.5T_\rmc$ and $0.25T_\rmc$, where
$T_\rmc = 0.019$.
In the paramagnetic phase at $T=1.05T_\rmc$,
we see a two-peak structure in $\sigma(\omega)$
 at $\omega \sim 0$ and $\omega = 2J$, and a gap in between.
As temperature is lowered below $T_\rmc$,
a transfer of the weight from the high-energy part $\omega \sim 2J$ to
the Drude part $\omega \sim 0$ is clearly observed.
The structure of the spectra and its temperature dependence is
found to be
qualitatively unchanged at $J \simge 1$ in the carrier doped system.
The nature of $\sigma(\omega)$ agrees with the experiment in a
qualitative way.

The above result
 is  understood from the change of the quasi-particle
DOS, which is shown in the inset of Fig.~\ref{FigOC}.
In the paramagnetic phase at $T > T_\rmc$,
the DOS equally splits into the lower sub-band and the upper sub-band
at $\omega \sim \mp J$, which are the spectra for quasi-particles with
spins parallel and anti-parallel to the localized spins, respectively.
The origin of the peaks at $\omega\sim 0$ and $\omega\sim 2J$
in $\sigma(\omega)$ is the intra-band and inter-band
excitation processes, respectively.
We note here that
the fermi level is at the lower band in the case of hole doping.
Below $T_\rmc$, polarization of localized spins give rise to
the change in the probability that an electron on a site
has a spin parallel (anti-parallel) to the localized spin on the site.
Therefore, the weights of
the sub-bands change as shown
in the figure.
Amplitude of the optical conductivity is roughly
 proportional to the product of the
DOS of particle and hole excitations with the same spin species.
Then,  as the magnetization increases,
the weight of $\sigma(\omega)$ transfers to the low frequency part.

It has been made clear in this paper
 that the transport properties of the system
are intensively influenced by the variation of the magnetic structure
because itinerant electrons and localized spins are strongly coupled.
Resistivity
is universally scaled by the  magnetization.
Neither temperature nor external magnetic field are essential
parameters.
Transfer in the weight of the optical conductivity is due to the
change in the quasi-particle DOS by the magnetization of
localized spins.

In order to compare
  dynamical processes of the system with
experiments, one should also treat the
 motions of localized spins which are  coupled not only
to itinerant electrons but also to the environment.
Within the current framework, such
dynamics of localized spins is replaced by
the thermal average over spin fluctuations.
The error due to this treatment is proportional to $1/D$ because
it is a spatial correlation term, so it vanishes at $D\to\infty$.
This replacement is also  valid  in finite dimensions if
the characteristic time-scale of the process
is much longer than the thermal
relaxation time of the localized spin $\tau_{\rm s} \simge 1/T$.
 Therefore, the present approach for the
conductivity $\sigma(\omega)$
is justified  at $\omega \ll 1/\tau_{\rm s}$ in the above sense,
 but may give inaccurate results for $\omega \simge 1/\tau_{\rm s}$
if we compare with experiments.
We speculate that this is one of the reasons why the optical conductivity
spectra at $\omega \sim 2J$
is not reproduced quantitatively while the resistivity
is in a good agreement with experiments.
In any cases, the qualitative nature of the transfer in the weight
of $\sigma(\omega)$ seems to be unchanged; it is mainly
determined by the quasi-particle DOS of the initial and the final states.

To summarize, we have calculated the Kondo lattice model with classical
spins  on a  Bethe lattice in $D=\infty$.
Temperature dependence of the conductivity is calculated.
As a function of the spontaneous magnetization,
resistivity is scaled in the same form as in
the case of MR.
The result agrees with the experimental data not only qualitatively
but also in a quantitative way.
Anomalous behavior in the optical conductivity of
{\LSMO} is explained from the change in the quasi-particle DOS.
Due to the strong coupling between itinerant electrons and
localized spins, transport properties
are intensively influenced by the magnetization of the system.

The author would like to thank Y. Tokura, T. Arima, F.F.\ Assaad
and M. Imada
for fruitful discussions and comments.
The numerical calculation is partially performed on the
 FACOM VPP500 at the Supercomputer Center,
Inst.\ for Solid State Phys., Univ.\ of Tokyo.

\vfil

\begin{figcaption}

\item Resistivity as a function of
total spontaneous magnetization (circle).
Resistivity as a function of total field-induced
magnetization for semi-circular
(triangle) and Lorentzian (dashed curve) DOS are also shown.
The experimental data  measured by changing magnetic field (square)
and temperature (solid curve) are taken from ref.~\rcite{Tokura94}.
\label{FigMR}

\item Optical conductivity for $J=2$ and $x=0.2$.
Inset shows the DOS in an arbitrary unit for the up-spin quasi-particle.

The down-spin DOS  at $\omega$ is
equal to  the up-spin DOS at $-\omega$ from the symmetry.
Temperatures are $T= 1.05T_{\rmc}$ (solid curves),
$T= 0.5T_{\rmc}$ (dashed curves) and $T= 0.25T_{\rmc}$ (dotted curves).

\label{FigOC}

\end{figcaption}

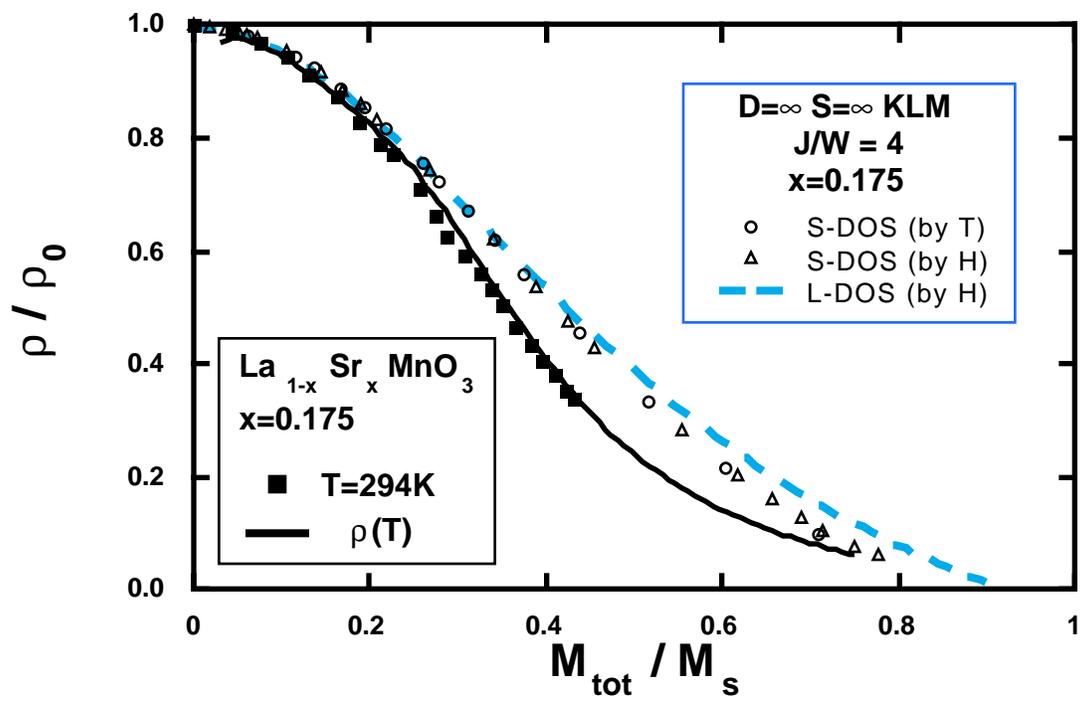
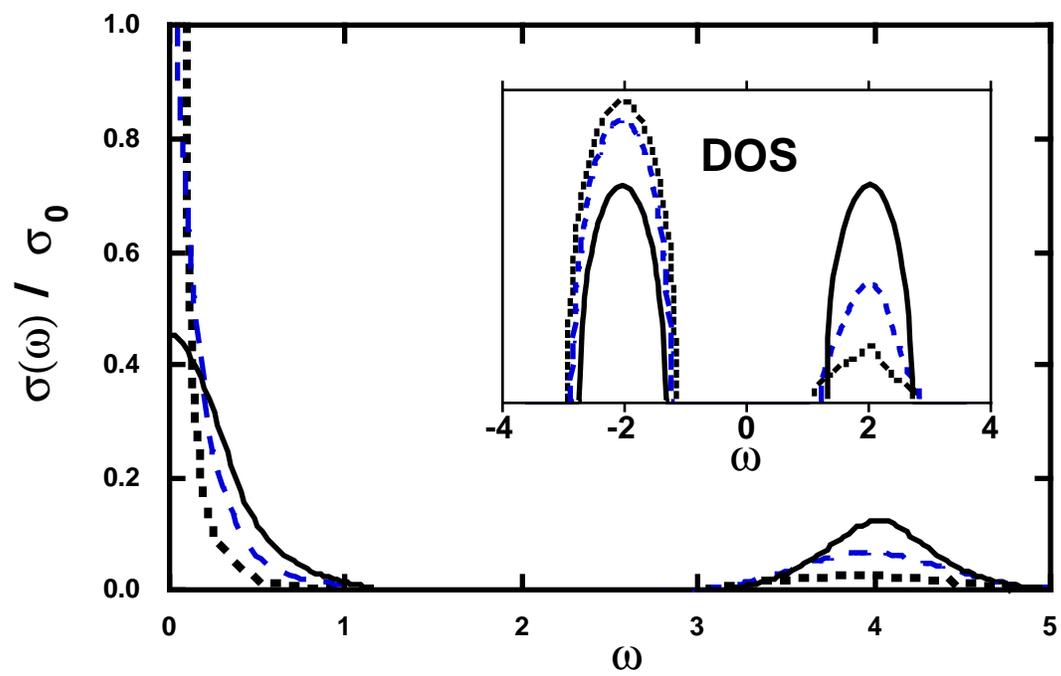
\end{document}